\documentclass[prb,aps,twocolumn]{revtex4}
\usepackage{graphicx}
\usepackage{epsfig}
\usepackage[matrix,frame,arrow]{xy}
\usepackage{amsmath}

\newcommand{\ket}[1]{\left\vert{#1}\right\rangle}
\newcommand{\qw}[1][-1]{\ar @{-} [0,#1]}
\newcommand{\qwx}[1][-1]{\ar @{-} [#1,0]}
\newcommand{\cw}[1][-1]{\ar @{=} [0,#1]}

\newcommand{\gate}[1]{*{\xy *+<.6em>{#1};p\save+LU;+RU **\dir{-}\restore\save+RU;+RD **\dir{-}\restore\save+RD;+LD **\dir{-}\restore\POS+LD;+LU **\dir{-}\endxy} \qw}
\newcommand{\meter}{\gate{\xy *!<0em,1.1em>h\cir<1.1em>{ur_dr},!U-<0em,.4em>;p+<.5em,.9em> **h\dir{-} \POS <-.6em,.4em> *{},<.6em,-.4em> *{} \endxy}}

\newcommand{\control}{*!<0em,.025em>-=-{\bullet}}

\newcommand{\ctrl}[1]{\control \qwx[#1] \qw}

\newcommand{\multigate}[2]{*+<1em,.9em>{\hphantom{#2}} \qw \POS[0,0].[#1,0];p !C *{#2},p \save+LU;+RU **\dir{-}\restore\save+RU;+RD **\dir{-}\restore\save+RD;+LD **\dir{-}\restore\save+LD;+LU **\dir{-}\restore}
\newcommand{\ghost}[1]{*+<1em,.9em>{\hphantom{#1}} \qw}

\newcommand{\rstick}[1]{*!L!<-.5em,0em>=<0em>{#1}}
\newcommand{\lstick}[1]{*!R!<.5em,0em>=<0em>{#1}}

\newcommand{\Qcircuit}[1][0em]{\xymatrix @*[o] @*=<#1>}

\begin{document}

\title{Arbitrary accuracy iterative phase estimation algorithm as a two qubit benchmark}

\author{Miroslav Dob\v{s}\'i\v{c}ek$^{1,2}$, G{\"o}ran Johansson$^1$,
Vitaly Shumeiko$^{1}$, and G{\"o}ran Wendin$^1$}

\affiliation{$^1$Microtechnology and Nanoscience, MC2, Chalmers,
S-412 96 G\"oteborg, Sweden \\
$^2$ Department of Computer Science and Engineering, FEE CTU, 121 35
Prague, Czech Republic}

\begin{abstract}
We discuss the implementation of an iterative quantum phase
estimation algorithm, with a single ancillary qubit. We suggest
using this algorithm as a benchmark for multi-qubit implementations.
Furthermore we describe in detail the smallest possible realization,
using only two qubits, and exemplify with a superconducting circuit.
We discuss the robustness of the algorithm in the presence of gate
errors, and show that 7 bits of precision is obtainable, even with
very limited gate accuracies.
\end{abstract}

\maketitle
Solid-state quantum computing is now entering the stage of
exploration of multi-qubit circuits. Coherent two-qubit coupling has
been experimentally realized for all major types of superconducting
qubits
\cite{Majer2005,Xu2005,terHaar2005,Grajcar2005,Ploeg2006,Niskanen2006,
Pashkin2003,Yamamoto2003,McDermott2005,Steffen2006,Jelle2006}, and
two-qubit gates have been demonstrated for charge
\cite{Yamamoto2003}, phase \cite{McDermott2005,Steffen2006} and flux
qubits \cite{Jelle2006}. The question then arises, what kind of
testbed application can be performed having at hand a very limited
amount of qubits?

Here we propose to employ the Phase Estimation Algorithm (PEA),
which can be implemented with just two qubits. Furthermore, we
suggest how to use this algorithm to characterize (benchmark) qubit
circuits. The PEA is an algorithm to determine the eigenvalue of a
unitary operator $\hat{U}$; it is closely related to the Quantum
Fourier Transform (QFT), which is a key element of many quantum
algorithms, e.g., Shor's factoring algorithm
\cite{ShorsFactoringAlgorithm} and in general Abelian Stabilizer
type of problems \cite{KitaevAlg}. The algorithm's relevance for
quantum simulations was noticed by Abrams and Lloyd
\cite{AbramsLloydPRL99}, and recently emphasized by Aspuru-Guzik
{\it et al.} \cite{AspuruGuzik2005} simulating quantum computation
of the lowest energy eigenvalue of several small molecules. It is
clear that the PEA will be one of the important algorithms in future
quantum information processing applications, and how accurately a
phase can be determined will be an important figure of merit for any
implementation.

The textbook \cite{Textbook} implementation of this algorithm
requires $n$ qubits representing the physical system in which
$\hat{U}$ operates, and $m$ ancillary qubits for the work register.
The number $m$ determines the algorithm's precision $1/2^m$, i.e the
number of accurate binary digits extracted.


There is also an alternative algorithm proposed by
Kitaev\cite{KitaevAlg}, where the Fourier transform is replaced with
a Hadamard transform. In implementing this algorithm to obtain a
precision of order $1/2^m$, it is possible to run either $\log{m}$
rounds (iterations) with $m$ ancillary qubits or $m\log(m)$ rounds
with only a single ancilla. The precision increases exponentially
with the number of rounds, but each round requires exponentially
many applications of $\hat{U}$, unless powers $U^{2^k}$ are
available by different means \cite{GiovannettiMetrology2006}.

Also the QFT-based PEA can be implemented in a multiround fashion,
using a single ancillary qubit, based on the semiclassical QFT
\cite{SemiQFT}. In this paper, we refer to this single ancilla QFT
based PEA as iterative PEA (IPEA). The iterative version of Kitaev's
algorithm is referred to as Kitaev's PEA. The relevance of the
iterative PEA as a viable alternative to the textbook version was
noticed by Mosca \& Ekert\cite{Mosca99} in the context of the hidden
subgroup problem, by Zalka\cite{Zalka98} for factoring, and by
Childs {\it et. al.} \cite{Childs2000} and Knill {\it et. al.}
\cite{Knill2007} in more physical contexts.

As long as the number of qubits is a limiting factor,
implementations of phase estimation with only a single ancillary
qubit will be of foremost importance. Thus, it is instructive to
compare the iterative PEA with Kitaev's PEA. In the IPEA scheme, the
bits of the phase are measured directly, without any need for
classical postprocessing. Moreover, each bit has to be measured only
once, compared to $\log(m)$ times. When the phase $\phi$ has a
binary expansion with no more than $m$ bits, the IPEA
deterministically extracts all bits, in contrast to Kitaev's PEA
which is always probabilistic. The IPEA is also optimal in the sense
that a full bit of information is gained in each
measurement\cite{vanDamPRL2007}.

Theoretically the accuracy of the algorithm is limited only by the
number of rounds, but in practice it will be limited by experimental
imperfections. Thus, the experimentally maximally obtainable
accuracy can serve as a benchmark for any multi-qubit
implementation.

For benchmarking purposes a setup is needed where the phase to be
measured can be set to an arbitrary value. We describe in detail
such an implementation in a system of two superconducting qubits.
Introducing gate noise, we also perform a robustness analysis,
indicating which gates are most critical, and we calculate the
number of repetitions needed as a function of noise levels.

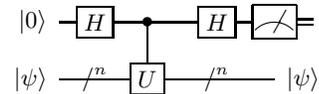
\begin{figure}[h!]
 \[
 \Qcircuit @R=1em @C=0.7em {
  \lstick{\ket{0}}    & \gate{H} & \ctrl{1} & \qw & \gate{H}
    & \meter &  \cw \\
  \lstick{\ket{\psi}} &  {/^n} \qw      & \gate{U} & \qw
    &  {/^n} \qw    & \rstick{\ket{\psi}} \qw }
 \]
 \caption{Naive implementation of the phase estimation algorithm.}
 \label{simplest_circuit_fig}
\end{figure}

{\em Iterative PEA.} We now describe the IPEA briefly, but still in
some detail, in order to make the robustness analysis clear. The
most straightforward approach for phase estimation is shown in
Fig.~\ref{simplest_circuit_fig}.
The upper line is the ancillary qubit which is measured, and the
lower line describes the qubits representing the physical system in
which $\hat{U}$ operates. Initially the ancillary qubit is set to
$\ket{0}$ and the lower line register to an eigenstate $\ket{\Psi}$
of the operator $\hat{U}$ with eigenvalue $e^{i2\pi \phi}$. Right
before the measurement the system state is
$\frac{1}{2}\left[\left(1+e^{i2\pi
\phi}\right)\ket{0}+\left(1-e^{i2\pi
\phi}\right)\ket{1}\right]\ket{\Psi}$, giving the probability
$P_0=\cos^2{(\pi\phi)}$ to measure $"0"$. By repeating this
procedure $N$ times, $P_0$ can be determined to an accuracy of
$1/\sqrt{N}$. Thus, one needs to go through at least $N \sim 2^{2m}$
independent rounds to obtain $m$ accurate binary digits of $\phi$.
The number of rounds corresponds to the number of measurements since
each round is terminated with a measurement.

Kitaev's PEA allows the number of rounds and consequently the number
of measurements to be drastically reduced, with the assumption that
the controlled-$\hat{U}^{2^k}$ gates are available \cite{KitaevAlg}.
For each $k$, $1 \leq k \leq m$, the controlled-$\hat{U}^{2^{k-1}}$
gate is used to prepare an ancillary qubit in the state
$\frac{1}{\sqrt{2}}\left(\ket{0}+e^ {i 2 \pi \left( 2^{k-1}  \phi
\right)} \ket{1} \right)$. After a number of repetitions, the ratio
of resulting zeros and ones is used as an estimate for the
fractional part of $2^{k-1}\phi$. A classical algorithm with
polynomial runtime is then used to assemble $\phi$ from the
fractional parts. The whole algorithm performs estimation of $\phi$
with precision $1/2^{m+2}$ and error probability $\leq \varepsilon$
after $O(m \log (m/\varepsilon))$ measurements. The gate $\hat{U}$
is applied $O(2^m \log (m/\varepsilon) )$ times to create the powers
$\hat{U}^{2^k}$, which is nearly a quadratic improvement compared to
the naive version of the PEA.

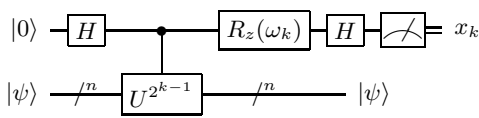
\begin{figure}[h!]
 \[
  \Qcircuit @R=1em @C=0.7em {
   \lstick{\ket{0}}    & \gate{H} & \ctrl{1}           &
      \gate{R_z(\omega_{k})}  & \gate{H} & \meter & \rstick{x_{k}} \cw \\
   \lstick{\ket{\psi}} &  {/^n} \qw      & \gate{U^{2^{k-1}}} &
   {/^n} \qw &
      \rstick{\ket{\psi}} \qw }
 \]
 \caption{The $k$th iteration of the iterative phase estimation
 algorithm.
 The feedback angle depends on the previously measured bits through
 $\omega_{k}=-2\pi(0.0x_{k+1}x_{k+2}\dots x_m)$, and $\omega_m=0$.
}
 \label{kick_back_circuit}
\end{figure}

The iterative PEA differs by the following modification of the above
described procedure: first less significant digits are evaluated and
then the obtained information improves the quantum part of the
search for more significant digits. The information transfer is done
with an extra single qubit Z-rotation that is inserted into the
circuit, as shown in Fig.~\ref{kick_back_circuit}. Note that $k$ is
iterated backwards from $m$ to 1.

In order to derive the success probability for each bit being
determined correctly, we first assume the phase $\phi$ to have a
binary expansion with no more than $m$ bits, $\phi=(0.\phi_1\phi_2
\dots \phi_m 0 0 0\ldots)$. In the first iteration ($k=m$) a
controlled-$\hat{U}^{2^{m-1}}$ gate is applied, and the $m$th bit of
the expansion is measured. The probability to measure "0" is
$P_0=\cos^2{\left[\pi (0.\phi_m 00\ldots)\right]}$, which is unity
for $\phi_m=0$ and zero for $\phi_m=1$. Thus, the first bit $\phi_m$
is extracted deterministically. In the second iteration ($k=m-1$)
the measurement is performed on the $(m-1)$th bit. The phase of the
first qubit before the Z-rotation is $2\pi(
0.\phi_{m-1}\phi_m00\ldots)$, and performing a Z-rotation with angle
$\omega_{m-1}=-2\pi (0.0\phi_m)$, the measurement probability
becomes $P_0=\cos^2{\left[\pi (0.\phi_{m-1}00\ldots)\right]}$. Thus,
using feedback the second bit is also measured deterministically,
and generally using the feedback angle
$\omega_{k}=-2\pi(0.0\phi_{k+1}\phi_{k+2}\dots \phi_m)$ all $m$ bits
of $\phi$ are extracted deterministically.

Denoting the first $m$ bits of the binary expansion of the phase
$\phi$ as $\tilde{\phi}=0.\phi_1\phi_2 \dots \phi_m$, there is in
general a remainder $0 \leq \delta < 1$, defined by
$\phi=\tilde{\phi}+\delta 2^{-m}$. In this case, the probability to
measure $\phi_m$  is $\cos^2{(\pi \delta/2)}$. If $\phi_m$ was
measured correctly, the probability to measure $\phi_{m-1}$ in the
second iteration is $\cos^2{(\pi \delta/4)}$, and so on. Thus, the
conditional probability $P_k$ for each bit to be measured correctly
is $P_k=\cos^2(\pi \, 2^{k-m-1} \delta)$, and the overall
probability for the algorithm to extract $\tilde{\phi}$ is
\begin{equation}
P(\delta)= \prod_{k=1}^{m} P_k =
 \frac{\sin^2{(\pi \, \delta)}}{2^{2 m} \sin^2{(\pi 2^{-m}\delta)}},
 \label{SuccessProbEq}
\end{equation}
which is the same outcome probability as the textbook phase
estimation, based on the QFT \cite{Textbook}. For $\delta \leq 1/2$
the best $m$-bit approximation to $\phi$ is indeed $\tilde{\phi}$,
while for $\delta
>1/2$ rounding up to $\tilde{\phi}+2^{-m}$ is better.
The probability to extract $\tilde{\phi}+2^{-m}$ is $P(1-\delta)$.
The success probability $P(\delta)$ decreases monotonically for
increasing $m$. In the limit $m\rightarrow\infty$, we find the lower
bound for the probability to extract the best rounded approximation
to $\phi$ as $P(1/2)=4/\pi^2$. Best rounding implies an error
smaller than $2^{-(m+1)}$, while an accuracy of $2^{-m}$ implies
that we accept both answers $\tilde{\phi}+2^{-m}$ and
$\tilde{\phi}$. The success probability is then
$P(\delta)+P(1-\delta)$, with a lower bound of $8/\pi^2$. In
conclusion, iterative PEA determines the phase with accuracy $1/2^m$
and with an error probability $\varepsilon < 1-8/\pi^2$, which is
{\em independent} of $m$.

Success probability amplification for the textbook PEA was discussed
in Ref.~\onlinecite{Textbook}. When the algorithm is executed with
$m'=m+\log{(2+1/2\varepsilon)}$ ancillas, the estimate is accurate
to $m$ bits, with probability at least $1-\varepsilon$. A similar
approach can also be used in iterative PEA, by determining $m'$ bits
and keeping only the $m$ most significant bits. However,
implementing the $\hat{U}^{2^{k}}$ gate for large $k$ is the
algorithm's bottleneck in a realistically noisy environment, as
discussed below.

A different approach, avoiding this problem, is to repeat the
algorithm a number of times, choosing the most frequent result. In
natural ensemble systems, such as NMR, this is naturally exploited
with advantage\cite{JonesPRL1999}. However, for single systems such
as superconducting qubits, repetitions on all $m$ bits are
unnecessarily expensive. From Eq.~(\ref{SuccessProbEq}) it is clear
that the main contribution to the error probability comes from the
least significant bits.

Thus, we may lower the error probability significantly by repeating
the measurement of only the first few bits a limited number of
times, and using simple majority voting. It is clear from the
binomial distribution that the bitwise error probabilities decrease
exponentially with the number of repetitions. Also, because of the
feedback procedure the bare error probability $\sin^2(\pi \,
2^{k-m-1} \delta)$ already decreases exponentially with decreasing
$k$. Thus, one needs only $O\left[\log^2{(1/\varepsilon)}\right]$
extra measurements to obtain an error probability smaller than
$\varepsilon$, independently of $m$.

{\em Benchmark circuit.} The minimal system for implementing the
iterative PEA is a two qubit system, where one qubit is a read-out
ancilla, and the second qubit represents a physical system. From the
work of Barenco {\it et al.} \cite{Barenco-Elementary-Gates} we know
an explicit construction of any controlled-$\hat{U}$ gate, where
$\hat{U}$ is an arbitrary single qubit gate. This construct involves
three single qubit gates and two controlled-NOT (CNOT) gates.

For benchmarking purposes we propose to use the very simple
Z-rotation operator
\begin{equation}
 \hat{U}= \left( \begin{array}{cc}
           e^{-i\alpha} &  0 \\
           0            &  e^{i\alpha}
          \end{array} \right),
          \label{ZrotationOperator}
\end{equation}
where $\alpha$ is an arbitrary rotation angle.

\begin{figure}[h!]
\[         \Qcircuit @R=.5em @C=0.4em {
                 \lstick{\ket{0}}
                 & \gate{R_x(\frac{\pi}{2})}
                 & \multigate{1}{ZZ\left(\alpha 2^{k-1}\right)}
                 & \qw
                 & \gate{R_z(-\omega_k)}
                 & \gate{R_x(-\frac{\pi}{2})}
                 & \meter
                 & \rstick{x_k} \cw \\
             \lstick{\ket{0} } 
                 & \qw
                 & \ghost{ZZ\left(\alpha 2^{k-1}\right)}
                 & \qw
                 & \rstick{\ket{0} } \qw 
         }
 \]
 \caption{A gate sequence implementing the $k$-th step of the
 iterative phase estimation algorithm, on a two qubit system
 using the entangling gate $ZZ(\alpha)$.
 }
 \label{TwoQubitImplementationFig}
\end{figure}
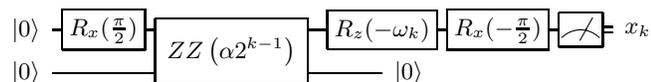

The advantages of this operator are 1) it is diagonal in the qubit
eigenbasis, thus the initial preparation of its eigenstate is
straightforward,
2) the phase to be measured is controlled directly, and 3)
controlled powers of this gate are generated by a single entangling
gate, $ZZ(\alpha) = {\rm
diag}\,(e^{-i\alpha},e^{i\alpha},e^{i\alpha},e^{-i\alpha})$; this
gate can be straightforwardly implemented by using most common
superconducting qubit coupling schemes.
As shown in Fig.~\ref{TwoQubitImplementationFig}, a step of the
iterative PEA is implemented using one ZZ-gate, and in addition only
three single qubit gates. The phase we are measuring in this case is
set by the coupling strength $\lambda$, rather than by the free
qubit energy that is the case using the general construction with
two CNOT gates, $\alpha 2^{k-1} = \lambda T$, $T$ is the pulse
duration.

Let us consider implementation of the ZZ-gate with superconducting
qubits in more detail. For superconducting charge and charge-phase
qubits operated at the charge degeneracy point, and physically
connected via a Josephson junction placed at the intersection of the
qubit loop-shaped electrodes, inductive interaction of persistent
currents circulating in the loops creates direct switchable
zz-coupling \cite{NoriPRB03,WendinShumeiko2005,LantzPRB04}. Thus the
implementation of the ZZ-gate is straightforward.

Furthermore, the ZZ-gate is a generic gate for the qubits coupled
via tunable linear oscillator: this gate is generated by applying a
composite dc-pulse sweeping through the qubit-cavity resonances as
shown in \cite{WallquistPRB06}.

For the permanent transverse coupling (xx-coupling in the qubit
eigenbasis) frequently discussed in the context of charge
\cite{Pashkin2003}, phase \cite{McDermott2005}, and flux
\cite{BertetPRB06} qubits, the ZZ-gate can be realized with dynamic
control schemes \cite{BertetPRB06,RigettiPRL05}. The parametric
coupling method \cite{BertetPRB06} suggests harmonic modulation of
the coupling strength $\lambda(t)$ with the two resonant frequencies
corresponding to the sum and the difference of the qubit energies.
This induces the Rabi rotations, $U_R^P$, and $U_R^S$, in the
parallel spin ($|00\rangle, |11\rangle$), and antiparallel spin
($|01\rangle, |10\rangle$) subspaces, respectively. The ZZ-gate is
then obtained, up to a single qubit $Z$-rotation, by applying the
Hadamard gates to both the qubits, ${\rm H} = {\rm H_1}\otimes{\rm
H_2}$, according to the scheme, $ZZ(\alpha) = {\rm
H}\,[U_R^P(\alpha)\oplus U_R^A(\alpha)]\,{\rm H}$.

The method \cite{RigettiPRL05} is similar although more time
consuming. In this case, the resonant rf-pulses are simultaneously
applied to both the qubits, inducing Rabi rotations $U_{R}$. The
pulse amplitudes are set equal to the half difference between the
qubit energies. Such an operation produces the gate which is
equivalent to the rotation in the $|00\rangle, |11\rangle$ subspace,
$U_{R}(\alpha) = {\rm H}U_{R}^P(\alpha/4){\rm H}$. Rotation in the
$|01\rangle, |10\rangle$ subspace can be performed by first swapping
the states of one of the qubits, and then applying the same pulse,
$U_{R}^S(\alpha/4) = X_1{\rm H}U_{R}(\alpha){\rm H}X_1$. Thus the
ZZ-gate is achieved applying the Rabi pulses twice, the full
operation sequence taking the form, $ZZ(\alpha) =
U_{R}(4\alpha)\,Z_1\,U_{R}(4\alpha)\,Z_1$.

\begin{figure}[h!]
\includegraphics[width=8cm]{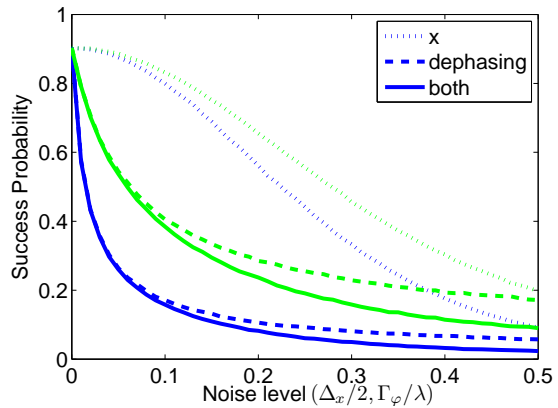}
\caption{The success probability of the IPEA to correctly determine
the phase $\alpha$, with precision better than $2^{-5}$ (green) and
$2^{-7}$ (blue) as a function of the noise level. The three cases of
pure $X$-gate errors (dotted), pure dephasing (dashed), and both
types of noise acting simultaneously (full) are considered. The
simulation was averaged over $\alpha$ evenly distributed in the
range $-\pi \leq \alpha < \pi$.} \label{gate_sensitivity_plot}
\end{figure}

{\em Robustness analysis.} There are numerous imaginable sources of
error, in all parts of the algorithm from initialization via gate
manipulation to readout. With our setup initialization will probably
be accurate, but the gates will certainly suffer from imperfections
due to environmental noise. First we consider the effect of pure
dephasing with rate $\Gamma_\varphi$, which eventually will limit
the accuracy of the $ZZ(\alpha 2^k)$ gate. In addition, we consider
imperfect $x$-rotations of the form $R_x(\pm\pi/2+\delta_x)$, where
$\delta_x$ is a normally distributed random angle with variance
$\Delta_x$. These errors modify the probability of measuring the
correct value of the $k$th bit into
\begin{equation}
P'_{k}=\frac{1}{2}\left[1+ e^{-\Delta_x^2 - |\alpha| 2^k
\Gamma_\varphi/\lambda } \cos{\left(\pi 2^{k-m} \delta \right)}
\right] . \label{SuccProbWithNoise}
\end{equation}

In Fig.~\ref{gate_sensitivity_plot} the algorithm's success
probability, as a function of the dephasing rate and variance
$\Delta_x$ is shown for $m=5$ and $m=7$. The algorithm is rather
robust against $x$-rotation errors, while being much more sensitive
to dephasing, which is evident from the exponentially growing factor
$2^k\alpha$ in the exponent of Eq.~(\ref{SuccProbWithNoise}).

\begin{figure}[t!]
\includegraphics[width=7.5cm]{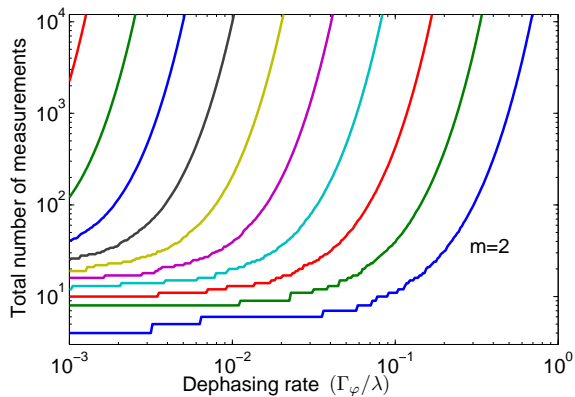}
\caption{The total number of measurements needed to obtain the phase
$\alpha$ with precision better than $2^{-m}$ ($2\leq m \leq 11$),
with an error probability $\varepsilon < 0.05$. }
\label{num_of_measurements_plot}
\end{figure}

As discussed below Eq.~(\ref{SuccessProbEq}), the success
probability can be improved by repeated measurements of each bit. To
achieve an overall success probability of $1-\varepsilon$, we need
to increase the per bit success probability to $(1-\varepsilon/m)$,
using $N_k$ repeated measurements. For $P'_k$ close to 1/2, many
repetitions are needed, and the binomial distribution approaches the
normal distribution giving,
\begin{equation}
N_k=\frac{1}{8}\left(\frac{erf^{-1}(1-\frac{2\varepsilon}{m})}{P'_k-\frac{1}{2}}\right)^2,
\end{equation}
where $erf^{-1}$ is the inverse of the error function. Considering
the dominating effect of dephasing, we find that the number of
repetitions grows quickly with the desired number of bits $m$,
$N_{k} \propto e^{2|\alpha|2^m\Gamma_\varphi/\lambda}$. In
Fig.~\ref{num_of_measurements_plot} we plot the total number of
measurements $N_{tot}=\sum_k N_k$, needed to obtain $2 \leq m \leq
11$ bits of the phase $\alpha$, with an error probability
$\varepsilon < 0.05$. For a realistic dephasing rate of 1 to 10
percent of the qubit-qubit coupling ($0.01 < \Gamma_\varphi/\lambda
< 0.1$), between 5-8 binary digits of $\alpha$ can be extracted with
less than $10^4$ measurements.

In conclusion, we have described a benchmark implementation of the
iterative PEA on two superconducting qubits, and analyzed its
robustness towards dephasing and gate errors. The number of
extractable binary digits is mainly limited by the dephasing, and
for realistic parameters amounts to 5-8. We believe phase estimation
will be an essential part of future applications of quantum
computing, and propose that the number of accurate binary digits can
be used as a benchmark for multi-qubit implementations.

This work is supported by the European Commission through the
IST-015708 EuroSQIP integrated project and by the Swedish Research
Council.


\begin{thebibliography}{99}
\bibitem{Majer2005}
J. B. Majer {\it et al.},
Phys. Rev. Lett. {\bf94}, 090501 (2005).

\bibitem{Xu2005}
H. Xu {\it et al.},
Phys. Rev. Lett. {\bf94},  027003 (2005).

\bibitem{terHaar2005}
A. C. J.  ter Haar, PhD thesis, Delft University (2005).

\bibitem{Grajcar2005}
M. Grajcar {\it et al.},
Phys. Rev. Lett. {\bf96}, 047006 (2006).

\bibitem{Ploeg2006}
S. H. W. van der Ploeg {\it et al.},
(2006); cond-mat/0605588.

\bibitem{Niskanen2006}
A. O. Niskanen, K. Harrabi, F. Yoshihara, Y. Nakamura, and J. S.
Tsai, Phys. Rev. B {\bf74}, 220503(R) (2006).

\bibitem{Pashkin2003}
Yu. A. Pashkin, T. Yamamoto, O. Astafiev, Y. Nakamura, D. V. Averin,
and J. S. Tsai, Nature {\bf421}, 823 (2003).

\bibitem{Yamamoto2003}
T. Yamamoto, Yu. Pashkin, O. Astafiev, Y. Nakamura, and J. S. Tsai,
Nature {\bf 425}, 941 (2003).

\bibitem{McDermott2005}
R. McDermott {\it et al.},
Science {\bf307}, 1299 (2005).

\bibitem{Steffen2006}
M. Steffen {\it et al.},
Science {\bf313} 1423 (206).

\bibitem{Jelle2006}
J. Plantenberg, P. C. de Groot, C. J. P. M. Harmans, and J. E.
Mooij,  Nature {\bf447}, 836 (2007).

\bibitem{ShorsFactoringAlgorithm}
P. W. Shor, SIAM J. Comput. {\bf 26} 1484 (1997).

\bibitem{KitaevAlg}
A. Yu. Kitaev, Electronic Colloquium on Computational Complexity
{\bf 3}(3) (1996).

\bibitem{AbramsLloydPRL99}
D. S. Abrams and S. Lloyd, Phys. Rev. Lett. {\bf83}, 5162 (1999).

\bibitem{AspuruGuzik2005}
Al\'{a}n Aspuru-Guzik {\it et al.},
Science {\bf 309}, 1704 (2005).

\bibitem{Textbook}
R. Cleve, A. Ekert, C. Macchiavello and M. Mosca, Proc. R. Soc.
London A, {\bf 454}, 339-354 (1998); Michael A. Nielsen and Isaac L.
Chuang, {\it Quantum Computation and Quantum Information}, Cambridge
University Press (2001).

\bibitem{GiovannettiMetrology2006}
V. Giovannetti, S. Lloyd, and L. Maccone,
Phys. Rev. Lett. {\bf 96}, 010401 (2006).

\bibitem{SemiQFT}
R. B. Griffiths and C.-S. Niu, Phys. Rev. Lett. {\bf76}, 3228
(1996).

\bibitem{Mosca99}
M. Mosca and A. Ekert,
Lecture Notes in Computer Science {\bf 1509}, 174 (1999).

\bibitem{Zalka98}
C. Zalka, arXiv:quant-ph/9806084.




\bibitem{Childs2000}
A. M. Childs, J. Preskill, J. Renes, J. Mod. Opt. {\bf 47}, 155-176
(2000);

\bibitem{Knill2007}
E. Knill, G. Ortiz, R. D. Somma, Phys. Rev. A {\bf 75}, 012328
(2007).

\bibitem{vanDamPRL2007}
W. van Dam, G. M. D'Ariano, A. Ekert, C. Macchiavello, and M. Mosca,
Phys. Rev. Lett. {\bf 98}, 090501 (2007).

\bibitem{JonesPRL1999}
J. A. Jones and M. Mosca,
Phys. Rev. Lett. {\bf 83}, 1050 (1999).


\bibitem{Barenco-Elementary-Gates}
A. Barenco {\it et al.},
Phys. Rev. A, {\bf 52}, 3457 (1995).

\bibitem{NoriPRB03}
J. Q. You and F. Nori, Phys. Rev. B {\bf 68}, 064509 (2003).

\bibitem{WendinShumeiko2005} G. Wendin and V. S. Shumeiko,
in {\it Handbook of Theor. and Comp. Nanotech.}, eds. M. Rieth and
W. Schommers, ASP, Los Angeles, (2006), Vol. 3, pp. 223-309.

\bibitem{LantzPRB04}
J. Lantz, M. Wallquist, V. S. Shumeiko, and G. Wendin, Phys. Rev. B
{\bf 70}, 140507(R) (2004).

\bibitem{WallquistPRB06}
M. Wallquist, V.S. Shumeiko, and G. Wendin, Phys. Rev. B {\bf74},
224506 (2006).

\bibitem{BertetPRB06}
P. Bertet, C. J. P. M. Harmans, and J. E. Mooij, Phys. Rev. B {\bf
73}, 064512 (2006).

\bibitem{RigettiPRL05}
C. Rigetti, A. Blais, and M. Devoret, Phys. Rev. Lett. {\bf 94},
240502 (2005).

\end{thebibliography}
\end{document}